\documentclass{article}

    \usepackage[final]{neurips_2022}

\usepackage[utf8]{inputenc} % allow utf-8 input
\usepackage[T1]{fontenc}    % use 8-bit T1 fonts
\usepackage{hyperref}       % hyperlinks
\usepackage{url}            % simple URL typesetting
\usepackage{booktabs}       % professional-quality tables
\usepackage{amsfonts}       % blackboard math symbols
\usepackage{nicefrac}       % compact symbols for 1/2, etc.
\usepackage{microtype}      % microtypography
\usepackage{xcolor}         % colors
\usepackage{amsmath}
\usepackage[linesnumbered,ruled]{algorithm2e}
\usepackage[flushleft]{threeparttable}
\usepackage{multirow}
\usepackage{wrapfig}
\usepackage{graphicx}
\usepackage{comment}
\usepackage{enumitem}

\title{Graph Coloring via Neural Networks for Haplotype Assembly and Viral Quasispecies Reconstruction}
\author{
  Hansheng Xue,\textsuperscript{\rm 1} Vaibhav Rajan,\textsuperscript{\rm 2} and Yu Lin\textsuperscript{\rm 1}\thanks{Corresponding author.} \\
  \textsuperscript{\rm 1}School of Computing, Australian National University, Canberra, Australia\\
  \textsuperscript{\rm 2}School of Computing, National University of Singapore, Singapore\\
  \texttt{\{hansheng.xue,yu.lin\}@anu.edu.au}, \texttt{vaibhav.rajan@nus.edu.sg}\\
}

\begin{document}

\maketitle

% \vspace{-15pt}
\begin{abstract}
\vspace{-5pt}
Understanding genetic variation, e.g., through mutations, in organisms is crucial to unravel their effects on the environment and human health.
A fundamental characterization can be obtained by solving the haplotype assembly problem, which yields the variation across multiple copies of chromosomes.
Variations among fast evolving viruses that lead to different strains (called quasispecies) are also deciphered with similar approaches.
% that may be two 
% %(e.g., in humans, called diploidy) 
% or more than two (called polyploidy), that is known as haplotype assembly and in the case of viruses, quasispecies reconstruction.
In both these cases, high-throughput sequencing technologies that provide oversampled mixtures of large noisy fragments (reads) of genomes, are used to infer constituent components (haplotypes or quasispecies).
%for haplotype assembly and quasispecies reconstruction.
The problem is harder for polyploid species where there are more than two copies of chromosomes.
State-of-the-art neural approaches to solve this NP-hard problem do not adequately model relations among the reads that are important for deconvolving the input signal.
We address this problem by developing a new method, called \texttt{NeurHap}, that combines graph representation learning with combinatorial optimization.
Our experiments demonstrate substantially better performance of \texttt{NeurHap}
%inference of haplotypes 
in real and synthetic datasets 
%and dramatic reduction of running times,
compared to competing approaches.

% Reconstructing haplotypes from the polyploid species and viral quasispecies plays a crucial role in genetics and has a profound impact on the environment and human health. Compared with diploid species, Few approaches focus on phasing polyploid species because the huge search space make the problem more challenging. 
% Existing methods focus on capturing the features of sequence with SNPs for each reads and neglect three core problems, a) MEC scores; b) implicit relations; c) SNPs sparsity. 
% To address these problems, we formulate the haplotype phasing problem as a graph coloring challenge on a MEC-based read-overlap graph. Then, we design a neural network search architecture to find a solution for read-overlap graph coloring. To address the ambiguity problem, we design a local refinement algorithm to further adjust colors for nodes. Extensive experiments on synthetic and real datasets demonstrate that our proposed \texttt{NeurHap} method significantly outperform state-of-the-art baselines for polyploid haplotypes phasing and viral quasispecies reconstruction.

\end{abstract}

% \vspace{-15pt}
\section{Introduction}
% \vspace{-5pt}
%\red{TODO: more general motivating paragraph}
Our genetic material is organized as sequences of DNA or RNA molecules (nucleotides) which form three-dimensional structures (chromosomes) within our cells.
Most organisms have multiple highly similar copies of chromosomes in their cells (e.g., humans have 2).
Variations in genetic sequences lead to the emergence of new species  %(e.g., virus strains) 
during evolution and are also 
known to be associated with many diseases (e.g., cancer).
% such as cancer.
There are many possible ways in which such variations can occur; the simplest among them is a {\it mutation} or a change in the nucleotide at a specific location in the DNA or RNA sequence.
A Single Nucleotide Polymorphism (SNP) refers to a mutation in at least one of the copies which renders the copies nonidentical at that point.
An ordered list of SNPs on a single chromosome is called a haplotype \citep{schwartz2010theory}.
Haplotypes provide a signature of genetic variability and thus inform us about disease susceptibilities and evolutionary patterns (e.g., of viruses).
These studies in turn pave the way for personalized medicine and effective drug development against viruses.

\begin{figure}[t]
\centering
% \vspace{-10pt}
\includegraphics[width=0.985\columnwidth]{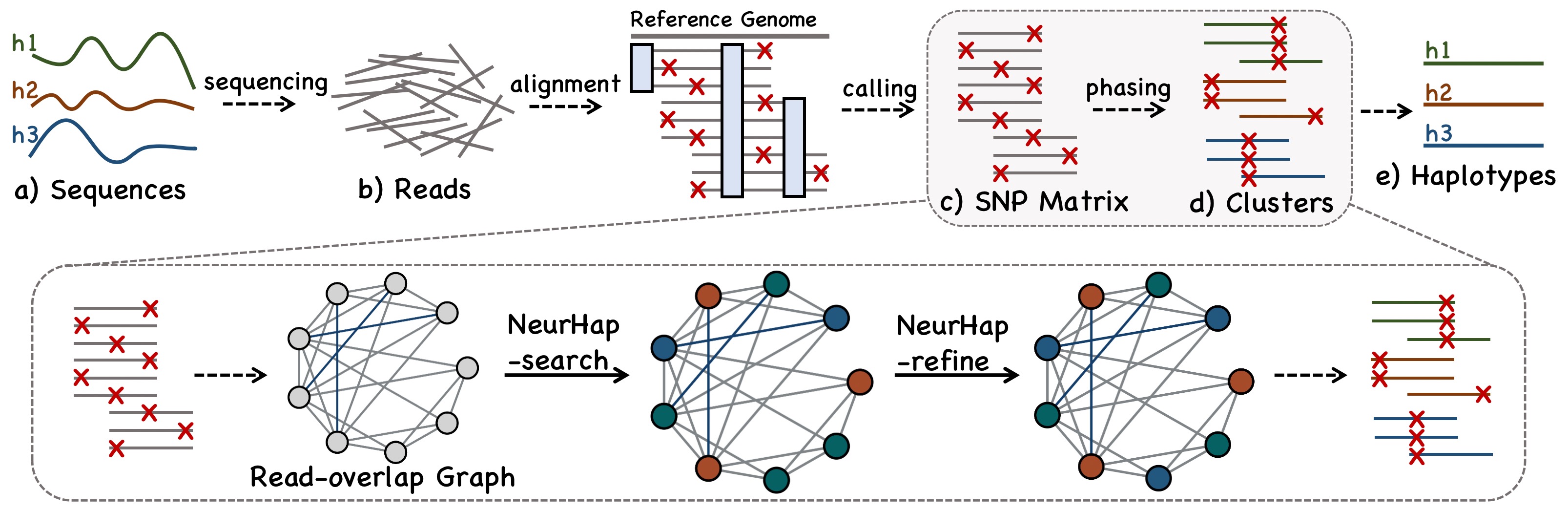}
% \vspace{-5pt}
\caption{The pipeline of reference-based polyploid haplotypes reconstruction and \texttt{NeuralHap}. Haplotype phasing is formulated as a graph coloring problem by constructing the Read-overlap graph.
\texttt{NeurHap} consists of \texttt{NeurHap}-search, a graph neural network to learn vertex representations and color assignments, and \texttt{NeurHap}-refine, a local refinement strategy to further adjust colors.}
\label{fig1}
% \vspace{-20pt}
\end{figure}

The problem of inferring haplotypes from 
high-throughput sequencing data is called haplotype assembly or phasing, and is done in multiple stages (see Figure \ref{fig1}).
Sequencing data yields multiple copies of short fragments of the entire genomic sequence (called reads, Figure \ref{fig1}b).
These reads are noisy due to sequencing errors and their short lengths may span across limited number of SNPs. This makes the problem of haplotype phasing challenging.
The reads are first aligned to a reference genome. 
This step indicates positions that are different across reads and thus infers the potential locations of SNPs. All other positions are discarded to obtain the SNP matrix (Figure \ref{fig1}c).
This matrix may be viewed as an oversampled mixture of noisy reads (restricted to SNPs).
Each mixture component represents a single haplotype and thus should have SNPs at the same locations.

In diploid species, containing two copies of chromosomes, there are two haplotypes to be inferred.
This problem has been studied extensively \citep{browning2011haplotype}.
In polyploid species, containing more than two copies of chromosomes (and thus more than two haplotypes), the problem is more challenging due to dramatic increase in search space~\citep{van2017evolutionary,abou2022towards,jablonski2021computational}.
In reconstruction of virus strains, called viral quasispecies, from viral populations, similar challenges arise. Moreover, 
unknown population sizes and imbalanced abundances pose additional difficulties \citep{jablonski2021computational}.

Existing approaches for haplotype phasing of polyploid species and viral quasispecies often group reads in the SNP matrix into clusters that correspond to different haplotypes, respectively. In an ideal case, all reads from the same cluster should be  consistent with respect to SNPs as they all belong to the same haplotype. 
In reality inconsistencies occur due to sequencing errors in reads.
Therefore, a minimum error correction (MEC) score~\citep{lippert2002algorithmic} is used to measure the discrepancy between the consensus haplotypes and their associated reads within each cluster (see Figure 1e). It is NP-hard to optimize the MEC score~\citep{zhang2006minimum}, and a number of combinatorial optimization heuristics have been proposed to approximate the optimal MEC score~\citep{zhang2020unzipping}.

More recently, the first neural network-based learning framework, named GAEseq~\citep{Ke2020GAEseq}, was proposed to phase haplotypes for polyploid species and viral quasispecies. CAECseq was later developed using a convolutional auto-encoder which captures spatial relationships between SNPs and enables clustering reads obtained from highly similar genomic regions ~\citep{Ke2020CAECseq}. Both GAEseq and CAECseq showed improved results compared to previous approaches.
A major limitation of
%these two methods is that
both CAECseq and GAEseq is that they cannot capture implicit relations among different reads. These methods have two independent steps (embedding and clustering) which makes the haplotype phasing results unstable. Besides, sparsity of the SNP matrix makes haplotype phasing for polyploids more challenging for these methods as well.

In this paper, we propose an  approach based on 
graph representation learning
for haplotype phasing of polyploid species and viral quasispecies. We formulate the haplotype phasing problem as a graph coloring problem, where the colors indicate haplotypes.
The graph is constructed from the SNP matrix where vertices are reads and two edge types are defined based on pairwise consistency and conflicts with respect to SNPs in the reads. 
Message passing-based neural networks are trained to minimize a loss designed to obtain a color assignment that
maximizes consistent edges and minimizes conflicting edges.
%based on the specified coloring constraints.
%to capture global topological information of the read-overlap graph. 
The network learns vertex representations and through them, an initial color assignment.
A local refinement strategy is then applied to adjust node colors in order to minimize MEC scores. 
%to fit the local structure of the read-overlap graph. 
Thus, in contrast to previous neural approaches that first learn representations and then cluster, our approach models the problem requirements in all steps.
As a result,  our model achieves better MEC scores, is more stable and also performs well on the challenging cases of polyploids and viral quasispecies.
In summary, our contributions are:
\begin{itemize}[noitemsep,topsep=0pt,labelindent=0em,leftmargin=*]
\item 
We provide a unique formulation of the haplotype phasing problem as a graph coloring problem, and develop an algorithm based on graph representation learning and combinatorial optimization.
\item
Our approach consists of \texttt{NeurHap}-search, a graph neural network to learn vertex representations and color assignments followed by \texttt{NeurHap}-refine, a local refinement strategy to adjust colors and optimize MEC scores. 
\item
Extensive experiments on synthetic and real datasets demonstrate that our new method \texttt{NeuralHap} significantly outperforms state-of-the-art phasing methods for both polyploid species and viral quasispecies.
\end{itemize}

\section{Related Work}
% \vspace{-5pt}
\textbf{Haplotype Phasing.} The aim of haplotype phasing of polyploid species and viral quasispecies
is to group reads into homogeneous clusters that corresponds to different haplotypes, respectively. The minimum error correction (MEC) score~\citep{lippert2002algorithmic} is introduced to measure the total discrepancy of reads in all clusters but is NP-hard to be optimised~\citep{zhang2006minimum}. Haplotype phasing for diploid species (i.e., reconstructing two haplotypes) has been extensively studied in the last two decades and a number of combinatorial optimization heuristics have been proposed to approximate the optimal MEC score, such as 
BNB~\citep{Wang2005HaplotypeRF},
HapCUT~\citep{Bansal2008HapCUTAE},
HASH~\citep{Bansal2008AnMA},
RefHap~\citep{Duitama2012FosmidbasedWG}
ProbHap~\citep{Kuleshov2014ProbabilisticSH},
HapCUT2~\citep{Edge2017HapCUT2RA} and others, and refer to ~\citep{zhang2020unzipping} for a recent review on phasing diploid species.

Haplotype phasing for polyploid species (i.e., reconstructing more than two haplotypes) becomes more computationally challenging as it requires a much larger search space compared to phasing two haplotypes for diploid species. A limited number of phasing methods work for polyploid species, e.g.,  
HapCompass~\citep{Aguiar2012HapCompassAF}, 
SDhaP~\citep{Das2015SDhaPHA}, 
H-PoP~\citep{Xie2016HPOP}, 
AltHap~\citep{hashemi2017AltHap}, 
refer to ~\citep{abou2022towards} for a recent review.
Haplotype phasing for viral quasispecies is very similar to the problem of phasing polyploid species. While haplotypes in polyploid species typically have uniform abundances, the different haplotypes (strains) in viral quasispecies may have varying abundances. Quite a few tools have also been proposed for haplotype phasing of viral quasispecies, 
such as 
ViSpA~\citep{Astrovskaya2011InferringVQ}, ShoRAH~\citep{Zagordi2011ShoRAHET},
QuRe~\citep{Prosperi2012QuReSF},
QuasiRecomb\citep{Tpfer2013ProbabilisticIO},
PredictHaplo~\citep{Prabhakaran2014HIVHI},
aBayesQR~\citep{Ahn2018aBayesQRAB},
TenSQR~\citep{Ahn2018ViralQR},
refer to ~\citep{jablonski2021computational} for a recent review.

More recently, deep learning models have been introduced into haplotype phasing for polyploid species and viral quasispecies. GAEseq~\citep{Ke2020GAEseq} uses a graph auto-encoder model on the constructed reads-SNPs bipartite network to model the relations between reads and SNPs. CAECseq~\citep{Ke2020CAECseq} uses a convolutional auto-encoder model to represent reads as low-dimensional features and then employs a clustering algorithm to group these reads. Note that GAEseq and CAECseq can be directly used to phase haplotypes for both polyploid species and viral quasispecies. Experimental results on both simulated and real datasets showed the superior results of GAEseq and CAECseq (in terms of MEC scores) compared to previous approaches for haplotype phasing for both polyploid species and viral quasispecies~\citep{Ke2020GAEseq,Ke2020CAECseq}. However, implicit relations among different reads have not been fully captured by GAEseq and CAECseq, especially when embedding and clustering are modelled separately and the SNP matrix is sparse.

\textbf{Neural Networks on Graphs.}
Most existing graph neural networks can be explained as a message-passing based graph learning model which recursively combines learned features/messages from their neighbors~\citep{Cui2019ASO,Cai2018ACS,Gilmer2017NeuralMP}. Popular methods include GCN~\citep{kipf2017GCN}, GraphSAGE~\citep{Hamilton2017graphsage}, GAT~\citep{velickovic2018gat} and GIN~\citep{xu2018GIN}. 
All these methods 
make the homophily assumption that similar nodes in the graph should be embedded close together. 
However, 
graph coloring aims to assign pairwise nodes with distinct colors for each edge of the graph, which is opposite to the homophily assumption. An intuitive way to integrate GNN models into the graph coloring challenge is to adjust the loss function, such as GNN-GCP~\citep{Lemos2019GraphCM}, RUN-CSP~\citep{Toenshoff2019RUNCSPUL}, and PI-GNN~\citep{Schuetz2022CombinatorialOW}.
% GNN-GCP~\citep{Lemos2019GraphCM} formulates a GNN-based solution for the graph coloring problem. RUN-CSP~\citep{Toenshoff2019RUNCSPUL} proposes a generic framework that uses a graph neural network model to solve binary constraint satisfaction problems.  PI-GNN~\citep{Schuetz2022CombinatorialOW} proposes a unifying framework incorporating insights from statistical physics that are broadly applicable to combinatorial optimization. 
% However, these models search for a solution over the conflict graph and cannot handle the consistent information simultaneously.
However, existing GNN-based graph coloring models cannot be implemented to the read-overlap graph directly because they cannot handle conflicting and consistent edges simultaneously.

\section{Methodology}
% \vspace{-5pt}
\textbf{Overview.}
In this paper, we propose the model of neural networks for graph coloring optimization to solve the haplotype reconstruction problem, called \texttt{NeurHap}. \texttt{NeurHap} mainly contains three steps, i) constructing the \textit{read-overlap graph}; ii) global coloring searching via iterative neural networks model, \texttt{NeurHap}-search; iii) local refinement to fine tune final coloring, \texttt{NeurHap}-refine. 

% \begin{figure}[t]
% \centering
% \includegraphics[width=0.6\columnwidth]{figures/figure_02.png}
% \caption{An illustrated example of constructing \textit{read-overlap graph}. }
% \label{fig2}
% \end{figure}

\begin{wrapfigure}{r}{0.4\textwidth}
% \vspace{-10pt}
  \centerline{\includegraphics[width=0.4\textwidth]{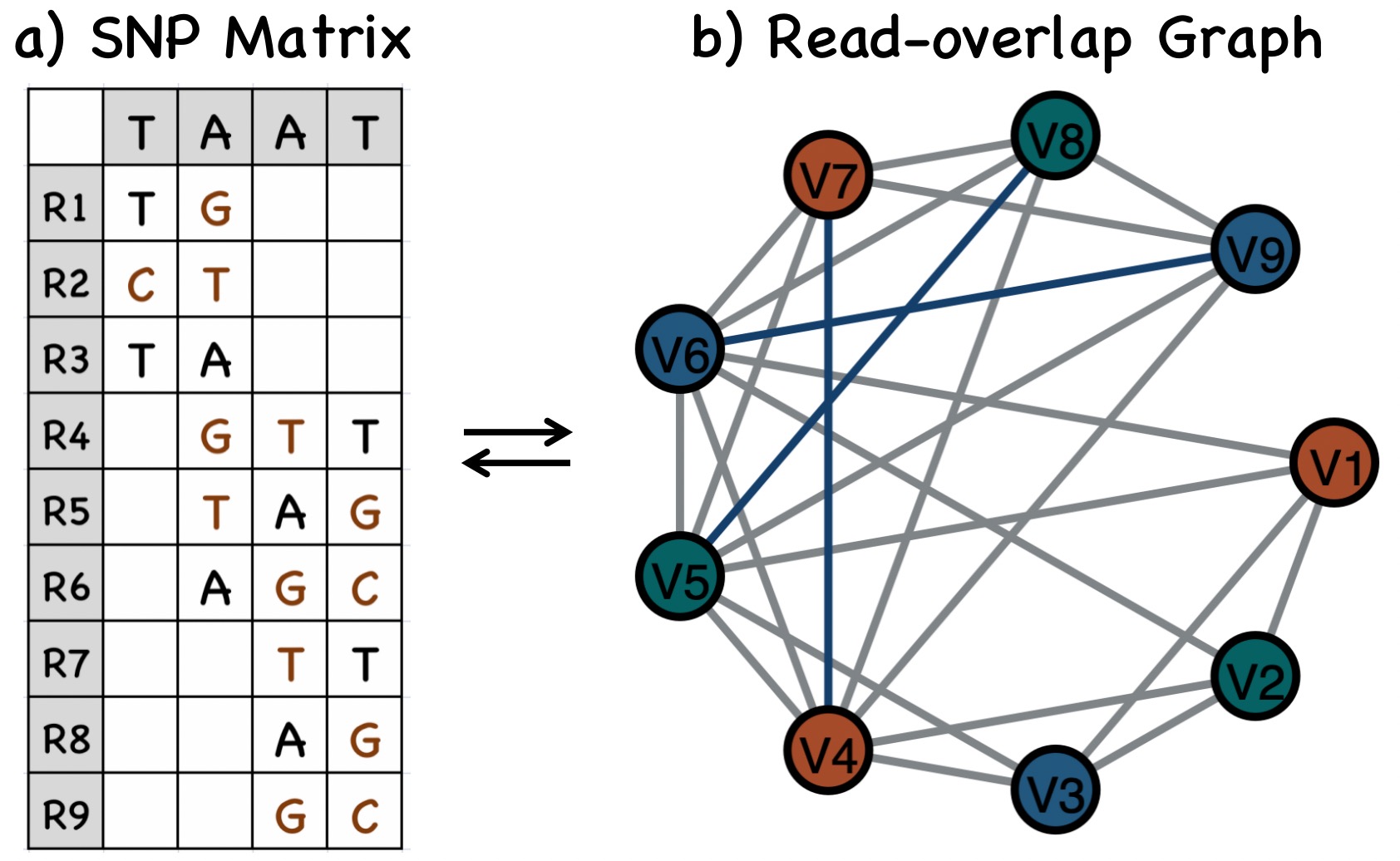}}
%   \vspace{-5pt}
  \caption{A toy example of constructing \textit{read-overlap graph} with conflict edges (in grey) and consistent edge (in blue).}
  \label{fig2}
% \vspace{-20pt}
\end{wrapfigure}

\textbf{Notations.}
Let $k$ be the number of haplotypes in a cell of polyploid species (aka. \textit{ploidy}) or the number of strains in viral quasispecies. For example, human and other mammals contain two haplotypes (diploid, \textit{k=2}); plants have more than 2 haplotypes (e.g., California redwood has six copies of each chromosome, hexaploid, \textit{k=6}). Viral quasispecies mixing 5 distinct HIV strains will have $k=5$.

Single nucleotide polymorphisms (SNPs) refer to positions where not all haplotypes have the same alleles. Given the alignment of reads to a reference genome, the SNP columns can be identified by removing the columns with identical alleles. The remaining alignment is referred to as a $m\times n$ SNP matrix $\mathcal{R}$ where $m$ denotes the number of reads and $n$ is the number of SNPs. The haplotype reconstruction aims to group $m$ reads into $k$ clusters , $\{C_1, C_2, \ldots, C_k\}$, that correspond to $k$ haplotypes, $\{\mathcal{H}_1, \mathcal{H}_2, \ldots, \mathcal{H}_k\}$, respectively. Once reads are grouped into clusters, the haplotype $\mathcal{H}_i$ can be reconstructed from reads in $C_i$ using a simple consensus voting. In the ideal case, all the reads from the cluster $C_i$ will be all consistent with $\mathcal{H}_i$. In reality, this is not the case and thus a minimum error correction (MEC) score~\citep{lippert2002algorithmic} is introduced to measure the discrepancy between the reads in the cluster $C_i$ and the consensus haplotypes $\mathcal{H}_i$ in all clusters. Given the grouping of reads into $k$ clusters $\{C_1, C_2, \ldots, C_k\}$, the corresponding MEC score can be computed as
\begin{equation}
\mathrm{MEC}(C_1,C_2,\ldots,C_k)=\sum_{i=1}^k\sum_{R_j \in C_i}HD(\mathcal{H}_i,R_j)
\label{eq01}
\end{equation}
where $HD(\cdot)$ is the Hamming distance function. 
Note that $HD(\mathcal{H}_i,R_j)$ for $R_j \in C_i$ can only be computed when we know all the reads in the cluster $C_i$ and use them to derive the consensus haplotype $\mathcal{H}_i$ of $C_i$. The main challenge in haplotype phasing is to find the grouping of reads into $\{C_1, C_2, \ldots, C_k\}$ such that the MEC score is minimized.

% In order to minimize the MEC score, reads from the same cluster should be as consistent as possible. 
Two reads are called \emph{overlapping} if they span over common SNP positions otherwise~\emph{non-overlapping}. 
Given any two reads, the relationship between them belongs to one of three cases, 
\emph{consistent}, \emph{conflict}, or \emph{ambiguous}. While the relationship between two non-overlapping reads is always \emph{ambiguous}, we further introduce two parameters $p$ and $q$ to define the relationship between two overlapping reads to account for sequencing errors and alignment ambiguity.  
Two overlapping reads are \emph{consistent} if they overlap at least $p$ positions and have the same alleles over all overlapping positions; are in \emph{conflict} if they differ on at least $q$ overlapping positions; and are \emph{ambiguous} otherwise.
The term `\emph{ambiguous}' means that there is not enough evidence to support that these two reads should belong to the same haplotype (`\emph{consistent}') or should belong to the different haplotypes (`\emph{conflict}'). 
For example, in Figure~\ref{fig2}, $R_4$ and $R_7$ are \emph{consistent}, $R_1$ and $R_2$ are in \emph{conflict}, and $R_1$ and $R_{4}$ are \emph{ambiguous}. 
In an ideal case, all the overlapping reads in the same cluster must be \emph{consistent}, i.e., if two reads are in \emph{conflict}, they must belong to different clusters. This observation naturally motivates us to build a read-overlap graph to model all reads as vertices and the important pairwise relationships (i.e., \emph{consistent} and \emph{conflict}) between overlapping reads as edges. Moreover, if we use $k$ colors to represent the $k$ clusters of reads, the problem of haplotype phasing is reduced to a graph coloring problem on the read-overlap graph. For example, in Figure~\ref{fig2}, the minimum MEC is achieved by grouping nine reads into three clusters, $C_1=\{R_1, R_4, R_7\}$, $C2=\{R_2, R_5, R_8\}$ and $C_3=\{R_3, R_6, R_9\}$, which correspond to three distinct colors on corresponding vertices in the read-overlap graph, respectively.

\subsection{Graph Coloring over the Read-overlap Graph.} 

The \textit{read-overlap graph} $\mathcal{G}=(\mathcal{V}, \mathcal{E}_{=}, \mathcal{E}_{\neq})$ is constructed in this step. Here, the vertex set $\mathcal{V}$ denotes all reads, the edge set $\mathcal{E}_{=}$ represents all pairwise consistent relationships between overlapping reads, and the edge set $\mathcal{E}_{\neq}$ refers to the pairwise conflict relationships between overlapping reads. 

Now we are ready to reduce the problem of haplotype phasing to the graph coloring problem on the read-overlap graph. Recall that the haplotype phasing problem aims to group reads into $k$ clusters such that reads from the same cluster are as consistent as possible. If we employ $c(v)$ to represent one out of $k$ colors assigned to a read $v$ (i.e., one of the $k$ clusters that $v$ belongs to), two reads $R_i$ and $R_j$ are in the same cluster if and only if two corresponding vertices $v_i$ and $v_j$ have the same color, i.e., $c(v_i) = c(v_j)$. Now the graph coloring problem needs to assign a color to $c(v)$ for every vertex $v\in\mathcal{V}$ to minimise the MEC score under the constraints that any two conflicting reads have two different colors and any two consistent reads have the same color.
\begin{equation}
\begin{aligned}
   & \min\ \mathrm{MEC}(c(v_1),c(v_2),\ldots,c(v_n))=\min \sum_{i=1}^k\sum_{c(v_j)=i}\mathrm{HD}(\mathcal{H}_i, R_j) \\
   & \mathrm{s.t.,}\ \left\{\begin{array}{rcl}
   \forall (v_i,v_j)\in\mathcal{E}_{\neq}, c(v_i)\neq c(v_j) \\
   \forall (v_i,v_j)\in\mathcal{E}_{=}, c(v_i)=c(v_j)
   \end{array}\right.
\end{aligned}
\label{eq02}
\end{equation}
Note that the above graph coloring problem is different from the classical graph coloring problem~\citep{pardalos1998graph} in combinatorial optimization. While all the edges are conflicting edges in the classical graph coloring problem, the above problem formulation in the equation~\ref{eq02} has constraints for both conflicting and consistent edges. In the following section, we will show how to model these constraints using neural networks.

\subsection{Network-based Global Search and Combinatorial Optimization-based Local Refinement}

\textbf{Satisfying Constraints.} As the vertices of the read-overlap graph need to be colored to satisfy the constraints in the equation~\ref{eq02}, we further reduce the graph coloring problem to a constraints satisfaction problem inspired by RUN-CSP~\citep{Toenshoff2019RUNCSPUL}. 
Graph neural networks (GNNs) 
are designed to follow
homophily constraints such that similar vertices in the graph are embedded close to each other (i.e., same colors). While this is true for consistent constraints in the read-overlap graph, the conflicting constraints impose heterozygous constraints, i.e., vertices connected by a conflicting edge should have very different embeddings (i.e., different colors). Therefore, given $k$ distinct colors, we introduce two $0/1$ metrics in $\mathbb{R}^{k\times k}$ for incorporating the above two different coloring constraints, $\mathcal{M}_{\neq}$ (for conflicting constraints) and $\mathcal{M}_{=}$ (for consistent constraints). The conflict-constraint matrix $\mathcal{M}_{\neq}$ denotes the binary conflict relationships among $k$ colors,
i.e., $\mathcal{M}_{\neq}(i,j)=1$ if $i\neq j$ and $0$ otherwise, for any $i, j \in\{1,...,k\}$. 
The consistent-constraint matrix $\mathcal{M}_{=}$ denotes the consistent relationships between $k$ colors, i.e., $\mathcal{M}_{=}(i,j)=1$ if $i=j$ and $0$ otherwise, for any $i, j \in\{1,...,k\}$. 

Assume that the coloring-assignment matrix $P\in\mathbb{R}^{|\mathcal{V}|\times k}$ is a matrix that represents the coloring assignment probability (over $k$ colors) for each vertex $v$ in its corresponding row $P(v)$. %We employ an unsupervised objective function to globally search for the optimal result. 
For any conflicting edge $(v_i,v_j)\in\mathcal{E}_{\neq}$ in the read-overlap graph, we aim to assign different colors to $v_i$ and $v_j$ and thus maximize the sum of joint-probabilities with different colors, i.e., $P(v_i)\mathcal{M}_{\neq}P(v_j)^{\mathsf{T}}$. Symmetrically, for any consistent edge $(v_i,v_j)\in\mathcal{E}_{=}$ in the read-overlap graph, we aim to assign the same color to $v_i$ and $v_j$ and thus maximize the sum of joint-probabilities with same colors, i.e., $P(v_i)\mathcal{M}_{=}P(v_j)^{\mathsf{T}}$. In summary, the unsupervised objective function can be formulated as follows:
\begin{equation}
    \mathcal{L}= -\frac{1}{|\mathcal{E}_{\neq}|}\sum_{(v_i,v_j)\in\mathcal{E}_{\neq}}\mathrm{log}(P(v_i)\mathcal{M}_{\neq}P(v_j)^{\mathsf{T}})-\lambda\cdot
    \frac{1}{|\mathcal{E}_{=}|}\sum_{(v_i,v_j)\in\mathcal{E}_{=}}\mathrm{log}(P(v_i)\mathcal{M}_{=}P(v_j)^{\mathsf{T}}).
\label{eq04}
\end{equation}
Here, $\lambda$ controls the importance of consistent constraints compared to conflicting constraints. We will now show how to use representation learnings to derive a coloring-assignment matrix $P$ that optimizes the above objective function.

\textbf{Global Search: \texttt{NeurHap}-search.} To derive a coloring-assignment matrix $P$, we propose to use an iterative message passing-based representation learning model to capture the structural information of the read-overlap graph. The message-passing learning model mainly contains three operation functions, message-learning, aggregation, and combine operator. 
Given trainable $d$-dimensional embeddings for every nodes $\{\mathrm{h}(v_1),\mathrm{h}(v_2),...,\mathrm{h}(v_n)\}, \mathrm{h}(v_i)\in\mathbb{R}^d,\ v_i\in\mathcal{V}$, which are initialized by randomly sampling from a uniform distribution, the message passing model can be formulated as: 
% \begin{equation}
% \sum_{v\in\mathcal{V}}\mathrm{agg}(\sum_{u\in N(v)}\mathrm{MLP}(\mathrm{h}_{v}||\mathrm{h}_{u}),\mathrm{h}_v)
% \end{equation}
% \begin{equation}
% \mathrm{h}_i = \mathrm{Agg}(\mathrm{msg}_i, \mathrm{h}_i),\ \ \mathrm{msg}_i = \sum_{j\in N(i)}\mathrm{MLP}(\mathrm{h}_{i}||\mathrm{h}_{j}).
% \end{equation}
% \begin{equation}
% \mathrm{h}_i = \mathrm{agg}(\mathrm{m}_i, \mathrm{h}_i),\ \ \mathrm{m}_i = \sum_{j\in N(i)}\mathrm{msg}(\mathrm{h}_{i},\mathrm{h}_{j}).
% \end{equation}
\begin{equation}
\begin{aligned}
 & \mathrm{h}(v_i) = \mathrm{COMBINE}(\mathrm{m}(v_i), \mathrm{h}(v_i)),\ \\
 & \mathrm{m}(v_i) = \mathrm{AGGREGATE}(\{\mathrm{MESSAGE}(\mathrm{h}(v_i),\mathrm{h}(v_j)): v_j\in N(v_i)\}).
\end{aligned}
\end{equation}
Here, $\mathrm{m}(v_i)$ is the learned messages from the neighbors of $v_i$, and $N(v_i)$ is the neighbors of $v_i$ with respect to conflict edges. $\mathrm{COMBINE}(\cdot)$ is a combine function and $\mathrm{AGGREGATE(\cdot)}$ denotes an aggregation function. To search for a simple model, we adopt a %most
recent message updater and mean operator
($\mathrm{m}(v_i)=\frac{1}{|N(v_i)|}\sum_{v_j\in N(v_i)}\mathrm{h}(v_j)$)
as combine and aggregate functions respectively. $\mathrm{MESSAGE(\cdot)}$ represents the learnable message function, e.g. $\mathrm{MESSAGE}(\mathrm{h}(v_i),\mathrm{h}(v_j))=\mathrm{MLP}(\mathrm{h}(v_i)||\mathrm{h}(v_j))$. Two linear layers with activation function (e.g., ReLU) are selected to construct the MLP layer. 
A simple linear decoder can be used to map the learned node embeddings to the probability of colors:  ${P}(v_i)=\mathrm{DEC}(\mathrm{h}(v_i))$. 
The message-passing model is iteratively trained 
for $t$ times in each epoch to generate reliable features for each node. 
The pseudocode for the global search process of \texttt{NeurHap} is as follows:

{\centering
\begin{minipage}{.95\linewidth}
\begin{algorithm}[H]
\SetAlgoLined
\KwData{SNP matrix $\mathcal{R}$; number of iteration $t$; number of polyploids $k$; dimension of hidden features $d$.}
\KwResult{Assignments $\mathcal{Y}$.}
$\mathcal{E}_{\neq},\mathcal{E}_{=}$ $\leftarrow$ Equation~\ref{eq01} \ \ \ \ \ \ \ \ //\ \texttt{Construct conflict  and consistent edge set}\\
 $\mathcal{M}_{\neq}(k)$, $\mathcal{M}_{=}(k)$ \ \ \ \ \ \ \ \ //\ \texttt{Initialize coloring constraints} \\
$\mathrm{h}$ $\leftarrow$ $\mathbb{R}^d{\sim[0,1)}$ \ \ \ \ \ \ \ \ //\ \texttt{Initialize by a uniform distribution}\\
\For{$e\ \mathrm{epochs}$}{
\For{$t\ \mathrm{iterations}$}{%\ \ \ //\ \texttt{Iterative Neural Networks Model}
 $\mathrm{\overline{m}}(v_j)=\mathrm{msg}(\mathrm{h}(v_i),\mathrm{h}(v_j))=\mathrm{MLP}(\mathrm{h}(v_i)||\mathrm{h}(v_j))$ \ \ \ \ \ \ //\ \texttt{Compute message from $h(v_i)$ and $h(v_j)$} \\
 $\mathrm{m}(v_i)=\mathrm{agg}(\mathrm{\overline{m}}(v_j): v_j\in N(v_i))$ \ \ \ \ //\ \texttt{Aggregate messages from neighbors of $v_i$} \\
 $\mathrm{h}(v_i) = \mathrm{comb}(\mathrm{m}(v_i), \mathrm{h}(v_i))$ \ //\ \texttt{Combine messages from $h(v_i)$ and $m(v_i)$} \\
}
$P$ $\leftarrow$ $\mathrm{dec}(\mathrm{h}(v_i))$ \ \ \ \ \ \ \ \ //\ \texttt{Compute coloring assignment probs} \\
$\mathcal{L}$ $\leftarrow$ Equation~\ref{eq04} \ \ \ \ \ \ \ \ //\ \texttt{Compute conflict loss} \\
$\mathcal{Y}$ $\leftarrow$ $P$ \ \ \ \ \ \ \ \ //\ \texttt{Compute coloring assignment} \\
}
\caption{The Global Search Algorithm \texttt{NeurHap}-search}
\end{algorithm}
\end{minipage}
\par
}

After optimizing the objective function with \texttt{NeurHap}, we can obtain an initial coloring assignment for vertices that satisfy the constraints in the equation~\ref{eq02} in the read-overlap graph. However, the objective function in equation~\ref{eq02} (i.e., the MEC score) may not be optimized as there may exist multiple coloring assignments that satisfy all constraints. Therefore, we
run an additional local refinement step
to further optimise the objective function in equation~\ref{eq02}.
%through a local refinement algorithm.

\textbf{Local refinement: \texttt{NeurHap}-refine.} 
%While \texttt{NeurHap}-search finds an initial coloring assignment that satisfy the conflict and consistent constraints, there is still room for improving the the objective function (the MEC score) in equation~\ref{eq02}. Here we propose a local refinement algorithm to address these issues, called \texttt{NeurHap}-refine. 
This step 
mainly searches for possible color adjustments of individual vertices given their associated conflicting and consistent constraints.
More specifically, if an individual vertex can be assigned a color different from its current color without violating any of the associated conflicting constraints with the neighboring vertices, the color is changed if a better MEC score is obtained by the change.
The refinement algorithm, \texttt{NeurHap}-refine, iteratively explores these possible color adjustments of individual vertices. 
%the MEC score can be further optimized. 
Refer to Appendix A.1 for the pseudocode.

\section{Experiments}
\textbf{Dataset.} 
To evaluate the proposed method \texttt{NeurHap}, we compare \texttt{NeurHap} with state-of-the-art baselines for both polyploid species and viral quasispecies. 
i) \textit{Polyploid species}: The Solanum Tuberosum is Tetraploid (\textit{k=4}) and the datasets of Solanum Tuberosum contains both simulated dataset \textbf{Sim-Potato} and real-world dataset \textbf{Real-Potato}, both downloaded from \citep{Ke2020CAECseq,Ke2020GAEseq}.
\textbf{Sim-Potato} contains 40 sub-datasets, which contains ten different samples sequenced at four distinct coverages (5X, 10X, 20X, and 30X). %The reference genome is selected as the Solanum Tuberosum chromosome \textit{5}~\citep{Xu2011GenomeSA} and Haplogenerator is used as the simulator to generate independent mutations according to a log-normal distribution~\citep{Motazedi2018ExploitingNS}. 
\textbf{Real-Potato} is the Chromosome \textit{5} capture-seq data of a small solanum tuberosum population available at NCBI (accession SRR6173308~\footnote{https://www.ncbi.nlm.nih.gov/sra/SRR6173308}). Ten samples are generated by randomly selecting ten genomic regions as the reference genome. 
ii) \textit{Viral Quasispecies}: Three viral quasispecies datasets are downloaded from SAVAGE\footnote{https://bitbucket.org/jbaaijens/savage-benchmarks}~\citep{Baaijens2017DeNA}, including the human immunodeficiency virus
(\textbf{5-strain HIV}, \textit{k=5}), the hepatitis C virus (\textbf{10-strain HCV}, \textit{k=10}), and the zika virus (\textbf{15-strain ZIKV}, \textit{k=15}). Ten samples are generated by randomly sampling from each of these three datasets.
%These datasets are generated using the true viral genomes from the NCBI database (SRR3332513
% \footnote{https://www.ncbi.nlm.nih.gov/sra/SRR3332513},
%, SRX396803 %\footnote{https://www.ncbi.nlm.nih.gov/sra/SRX396803}) 
%and the Illumina MiSeq reads are simulated used the software SimSeq~\citep{Benidt2015SimSeqAN}. We randomly select 10 samples from the download datasets. 
% For 5-strain HIV, we split the whole dataset into 12 samples. For 10-strain HCV and 15-strain ZIKV, we randomly selected 10 samples from the downloaded datasets. 
%
In this paper,  we use BWA-MEM~\citep{Li2013AligningSR} to align reads to the reference genome
and use the same tool described in CAECseq and GAEseq \citep{Ke2020CAECseq,Ke2020GAEseq} 
to derive the SNP matrix from the above alignment to ensure a fair comparison. 

\textbf{Baseline algorithms.} 
GAEseq~\citep{Ke2020GAEseq} and CAECseq~\citep{Ke2020CAECseq} 
are two state-of-the-art approaches that work 
on both haplotype assembly and viral quasispecies reconstruction.
We included two additional methods, H-PoP~\citep{Xie2016HPOP},  AltHap~\citep{hashemi2017AltHap}, that specifically work on haplotype assembly for polyploid species.
We also included one additional method, TenSQR~\citep{Ahn2018ViralQR}, 
that specifically works on viral quasispecies reconstruction. 
Many other specific methods are not included in this study 
because GAEseq~\citep{Ke2020GAEseq} and CAECseq~\citep{Ke2020CAECseq} 
have recently demonstrated their superior performance against other baselines
in both haplotype assembly and viral quasispecies reconstruction.

\textbf{Experimental setup.}
The minimum error correction (MEC) score, given in equation~\ref{eq01}, is adopted as the evaluation metric~\citep{Lippert2002MEC} for both haplotype assembly and viral quasispecies reconstruction.
% The MEC score is designed to measure the difference between reads and the reconstructed haplotypes at positions that are covered by the reads. 
%The detailed equation of the MEC is formulated at equation~\ref{eq01}.
Following the experimental setup in \citep{Ke2020CAECseq,Ke2020GAEseq}, all the algorithms 
run ten times on each input dataset and the lowest MEC score is reported. 
%For all baseline methods, we optimize their models with different parameters and report the best performance scores.
The initial number of polyploids $k$ is known: $k=4$ for both Sim-Potato and Real-Potato, $k=5$ for 5-strain HIV, $k=10$ for 10-strain HCV, and $k=15$ for 15-strain ZIKV. 
%Several important parameters are included in
The default settings of \texttt{NeurHap} hyperparameter are as follows.
The representation dimensions are all empirically set to be 32. 
The number of iteration $t$ in \texttt{NeurHap}-search is set to be 10 as default. 
The parameter $\lambda$ chooses 0.01 as the default value. 
The default values for parameters $p$ and $q$ are 3 and 5, respectively. 
% All codes, data and experimental settings for \texttt{NeurHap} model will be released after the double-blind review.
% The NeurHap model is freely available~\footnote{https://github.com/xuehansheng/NeurHap}.
The NeurHap model is freely available at \url{https://github.com/xuehansheng/NeurHap}.

\subsection{Performance on Polyploid Species data}

\begin{wraptable}{r}{0.6\textwidth}
\vspace{-15pt}
 \caption{Performance comparison on Sim-Potato data.}
  \centering 
  \small
  \setlength{\tabcolsep}{0.9mm}
  \begin{tabular}{c|c|c|c|c}
    \toprule
    Model & \#Cov 5X & \#Cov 10X & \#Cov 20X & \#Cov 30X \\
    \midrule
     H-PoP & 429.0\scriptsize$\pm$64.1 & 933.9\scriptsize$\pm$103.6 & 1782.2\scriptsize$\pm$161.8 & 2826.9\scriptsize$\pm$180.7 \\
     AltHap & 610.9\scriptsize$\pm$259.3 & 722.3\scriptsize$\pm$179.1 & 649.3\scriptsize$\pm$369.4 & 1148.2\scriptsize$\pm$509.9 \\
     GAEseq & 153.7\scriptsize$\pm$20.3 & 261.6\scriptsize$\pm$58.7 & 372.8\scriptsize$\pm$74.5 & 496.9\scriptsize$\pm$128.7 \\
    %  CAECseq & \textcolor{blue}{96.2\scriptsize$\pm$26.9} & \textcolor{blue}{141.4\scriptsize$\pm$40.7} & \textcolor{blue}{254.2\scriptsize$\pm$99.7} & \textcolor{blue}{372.9\scriptsize$\pm$148.9} \\
    CAECseq & \underline{96.2\scriptsize$\pm$26.9} & \underline{141.4\scriptsize$\pm$40.7} & \underline{254.2\scriptsize$\pm$99.7} & \underline{372.9\scriptsize$\pm$148.9} \\
    \cmidrule{1-5}
    %  NeurHap & \textcolor{violet}{\textbf{29.9\scriptsize$\pm$5.7}} & \textcolor{violet}{\textbf{51.9\scriptsize$\pm$8.2}} & \textcolor{violet}{\textbf{92.6\scriptsize$\pm$10.6}} & \textcolor{violet}{\textbf{142.0\scriptsize$\pm$23.6}} \\
    NeurHap & \textbf{29.9\scriptsize$\pm$5.7} & \textbf{51.9\scriptsize$\pm$8.2} & \textbf{92.6\scriptsize$\pm$10.6} & \textbf{142.0\scriptsize$\pm$23.6} \\
    % (Improv.) & \\
    \bottomrule
  \end{tabular}
  \label{tab01}
  \vspace{-10pt}
\end{wraptable}

Table~\ref{tab01} and ~\ref{tab02} show that \texttt{NeurHap} significantly outperforms state-of-the-art baselines, which achieving the lowest MEC scores on both the Sim-Potato and Real-Potato datasets. For Cov-5X of Sim-Potato, the MEC score obtained by \texttt{NeurHap} is 29.9 which is about 3x lower than the lowest scores achieved by baselines (96.2 for CAECseq). For Real-Potato, \texttt{NeurHap} also achieves the lowest MEC scores on all samples. The average MEC score achieved by \texttt{NeurHap} is 371.6 which is significantly lower than the second lowest MEC score obtained by CAECseq, 400. The gap between \texttt{NeurHap} and baselines demonstrates the superiority of our model in polyploid haplotype phasing.

\begin{table}[!htb]
% \vspace{-10pt}
 \caption{Performance comparison on Real-Potato data.}
  \centering 
  \small
  \begin{tabular}{c|ccccccccccc}
    \toprule
    Sample & \#1 & \#2 & \#3 & \#4 & \#5 & \#6 & \#7 & \#8 & \#9 & \#10 & Avg. \\
    \cmidrule{1-12}
    Reads & 240 & 389 & 274 & 115 & 141 & 398 & 295 & 284 & 489 & 449 & - \\
    SNPs & 294 & 238 & 83 & 23 & 176 & 198 & 456 & 424 & 236 & 410 & - \\
    \midrule
    \midrule
    H-PoP & 705 & 525 & 132 & 4 & 240 & 982 & 981 & 766 & 793 & 1413 & 654.1\scriptsize$\pm$435.6 \\
    AltHap & 746 & 572 & 192 & 9 & 299 & 1295 & 1021 & 982 & 811 & 1311 & 723.8\scriptsize$\pm$451.1 \\
    % H-PoP & 235 & 460 & 140 & 4 & \textcolor{blue}{168} & 917 & 571 & 613 & \textcolor{violet}{\textbf{534}} & 751 & 441.1\scriptsize$\pm$ 277.4 \\
    % AltHap & 516 & 557 & 241 & 11 & 342 & 1124 & 986 & 1238 & 947 & 1059 &  702.1\scriptsize$\pm$401.1 \\
    GAEseq & 231 & 406 & \underline{97} & \underline{2} & 180 & 873 & 558 & 441 & \underline{592} & 712 & 409.2\scriptsize$\pm$266.6 \\
    CAECseq & \underline{229} & \underline{393} & 103 & \textbf{1} & 172 & \underline{859} & \underline{522} & \underline{430} & 593 & \underline{698} & \underline{400.0\scriptsize$\pm$260.9} \\
    \cmidrule{1-12}
    NeurHap & \textbf{178} & \textbf{343} & \textbf{93} & \textbf{1} & \textbf{163} & \textbf{857} & \textbf{499} & \textbf{384} & \textbf{561} & \textbf{632} & \textbf{371.6\scriptsize$\pm$268.9} \\
    % (Improv.) & \\
    % \midrule
    % GAEseq & 275.7 & 415.0 & 175.1 & 38.5 & 230.5 & 963.6 & 580.3 & 512.7 & 640.5 & 787.0 & 461.9 \\
    % CAECseq & 284.1 & 435.6 & 121.2 & 4.7 & 209.7 & 916.5 & 594.0 & 472.8 & 648.2 & 798.6 & 448.5 \\
    % % \hline
    % % Greedy &  \\
    % % Tabucol & \\
    % % GNN-GCP & \\
    % % \midrule
    % NeurHap &  \\
    % (Improv.) & \\
    \bottomrule
  \end{tabular}
  \label{tab02}
%   \vspace{-10pt}
\end{table}

\subsection{Performance on Viral Quasispecies data}
Table~\ref{tab03} shows the results obtained by \texttt{NeurHap} and baselines for reconstructing viral quasispecies on three datasets respectively, 5-strain HIV, 10-strain HCV, and 15-strain ZIKV. 
In Table~\ref{tab03}, \texttt{NeurHap} significantly outperforms baselines on all 10 samples in these datasets. \texttt{NeurHap} achieves the lowest MEC score in 5-strain HIV (1371.4), which is about 160 lower than the MEC score obtained by CAECseq (1638.5). 
For 10-strain HCV data, \texttt{NeurHap} also achieves the lowest MEC score 1008.1 and the second lowest MEC score is 1144.3 obtained by TenSQR. 
With increase in the number of haplotypes (strains),
%two universal polyploid haplotype reconstruction methods 
performance of 
CAECseq and GAEseq deteriorates and while that of \texttt{NeurHap} improves. \texttt{NeurHap} significantly outperforms CAECseq and GAEseq on polyploid haplotypes.

\begin{table}[!htb]
% \vspace{-10pt}
 \caption{Performance comparison on three viral quasispecies datasets.}
  \centering 
  \small
  \setlength{\tabcolsep}{1.2mm}
  \begin{tabular}{c|c|ccccccccccc}
    \toprule
    Dataset & & \#1 & \#2 & \#3 & \#4 & \#5 & \#6 & \#7 & \#8 & \#9 & \#10 & Avg. \\
    \midrule
    \multirow{6}{*}{5-strain} & Reads & 967 & 961  & 951 & 961 & 966 & 969 & 962 & 965 & 955 & 971 & - \\
    \multirow{6}{*}{HIV} & SNPs & 1617 & 1685 & 1595 & 1605 & 1615 & 1660 & 1619 & 1622 & 1580 & 1653 & - \\
    % \midrule
    % \midrule
    \cmidrule{2-13}
    % AltHap & 5297 & 5237 & 6019 & 3996 & 6275 & 4473 & 4917 & 2960 & 4402 & 6287 & 3858 & 4714 & 4869.6\scriptsize$\pm$1020.6 \\
    & TenSQR & 1920 & 2324 & 1867 & 1896 & 2055 & 1793 & 2125 & 1754 & 1679 & 1757 & 1917.0\scriptsize$\pm$198.6 \\
    & GAEseq & 1981 & 1953 & 1678 & 1806 & 1905 & 2007 & 1819 & 1746 & 1702 & 1747 & 1834.4\scriptsize$\pm$119.6\\
    & CAECseq & \underline{1729} & \underline{1750} & \underline{1787} & \underline{1552} & \underline{1730} & \underline{1622} & \underline{1611} & \underline{1529} & \underline{1519} & \underline{1556} & \underline{1638.5\scriptsize$\pm$101.5} \\
    \cmidrule{2-13}
    & NeurHap & \textbf{1307} & \textbf{1525} & \textbf{1385} & \textbf{1265} & \textbf{1410} & \textbf{1382} & \textbf{1393} & \textbf{1323} & \textbf{1274} & \textbf{1450} & \textbf{1371.4\scriptsize$\pm$81.2}\\
    \midrule
    \midrule
    \multirow{6}{*}{10-strain} & Reads & 500 & 498 & 500 & 499 & 498 & 500 & 499 & 500 & 500 & 500 & - \\
    \multirow{6}{*}{HCV} & SNPs & 1770 & 1712 & 1794 & 1749 & 1741 & 1759 & 1786 & 1765 & 1743 & 1808 & - \\
    % \midrule
    \cmidrule{2-13}
    & TenSQR & \underline{1081} & \underline{1037} & \underline{1106} & \underline{960} & \underline{1115} & \underline{1015} & \underline{1365} & 1293 & 1396 & \underline{1075} & \underline{1144.3\scriptsize$\pm$151.8} \\
    & GAEseq & 1270 & 1121 & 1301 & 1171 & 1245 & 1152 & 1371 & \underline{1105} & \underline{1152} & 1200 & 1208.8\scriptsize$\pm$85.8 \\
    & CAECseq & 1490 & 1616 & 1347 & 1675 & 1475 & 1405 & 1563 & 1413 & 1436 & 1554 & 1497.4\scriptsize$\pm$103.1 \\
    \cmidrule{2-13}
    & NeurHap & \textbf{1029} & \textbf{990} & \textbf{1097} & \textbf{956} & \textbf{1012} & \textbf{899} & \textbf{1014} & \textbf{1008} & \textbf{1079} & \textbf{997} & \textbf{1008.1\scriptsize$\pm$56.3} \\
    \midrule
    \midrule
    \multirow{6}{*}{15-strain} & Reads & 500 & 500 & 500 & 500 & 500 & 499 & 498 & 497 & 499 & 500 & - \\
    \multirow{6}{*}{ZIKV} & SNPs & 2384 & 2358 & 2385 & 2360 & 2386 & 2383 & 2375 & 2373 & 2353 & 2353 & - \\
    \cmidrule{2-13}
    & TenSQR & \underline{941} & \underline{794} & \underline{859} & \underline{869} & \underline{950} & \underline{856} & \underline{848} & \underline{789} & \underline{849} & \underline{758} & \underline{851.3\scriptsize$\pm$61.5} \\
    & GAEseq & 1470 & 1585 & 1515 & 1590 & 1713 & 1363 & 1523 & 1348 & 1618 & 1393 & 1511.8\scriptsize$\pm$119.4 \\
    & CAECseq & 2344 & 2248 & 2427 & 2338 & 2454 & 2406 & 2378 & 2496 & 2292 & 2309 & 2369.2\scriptsize$\pm$77.4 \\
    \cmidrule{2-13}
    & NeurHap & \textbf{718} & \textbf{655} & \textbf{752} & \textbf{721} & \textbf{862} & \textbf{658} & \textbf{694} & \textbf{622} & \textbf{666} & \textbf{675} & \textbf{702.3\scriptsize$\pm$67.8}\\
    \bottomrule
  \end{tabular}
  \label{tab03}
%   \vspace{-10pt}
\end{table}

\subsection{Visualization}
\begin{figure}[t]
\centering
\includegraphics[width=0.95\columnwidth]{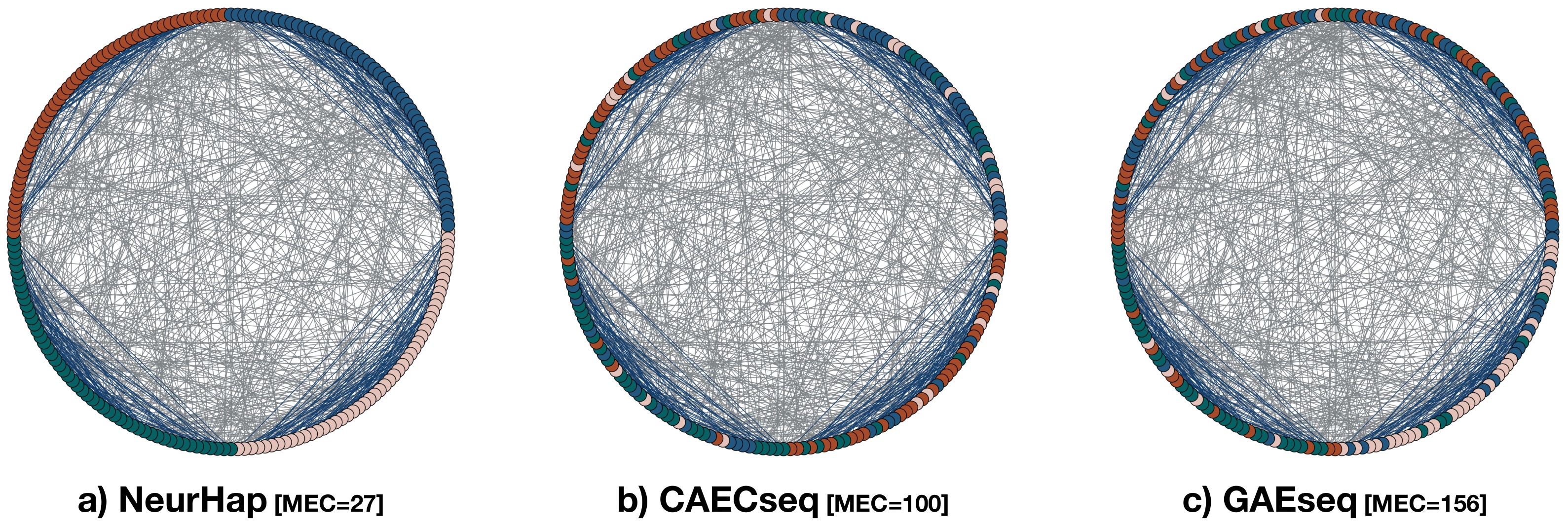}
% \vspace{-5pt}
\caption{The visualization of \texttt{NeurHap}, CAECseq, and GAEseq on Sim-Potato-5X-Sample1 data.}
\label{fig4}
% \vspace{-5pt}
\end{figure}

To better understand the phasing results, we use python-iGraph package to visualize the read-overlap graph of Sim-Potato-5X dataset with clustering results from \texttt{NeurHap}, CAECseq, and GAEseq (see Figure~\ref{fig4}). Different colors denote distinct haplotypes (the number of haplotypes for Sim-Potato is 4). Grey edges are conflicting edges and blue edges are consistent edges in the read-overlap graph. 
The color of nodes in Figure~\ref{fig4} are derived from the clusters constructed by different models, i.e., each color indicates a cluster of reads that are inferred to come from the same haplotype. 
In Figure~\ref{fig4} b) and c), 89 and 133 conflicting edges are violated (i.e., connecting two vertices with the same color) for CAECseq and GAEseq, respectively, while none of the conflicting edges are violated for \texttt{NeurHap}. 
\texttt{NeurHap} derives a coloring assignment that is most consistent with the conflicting and consistent edges in the read-overlap graph.

\begin{figure}[t]
\centering
\includegraphics[width=0.95\columnwidth]{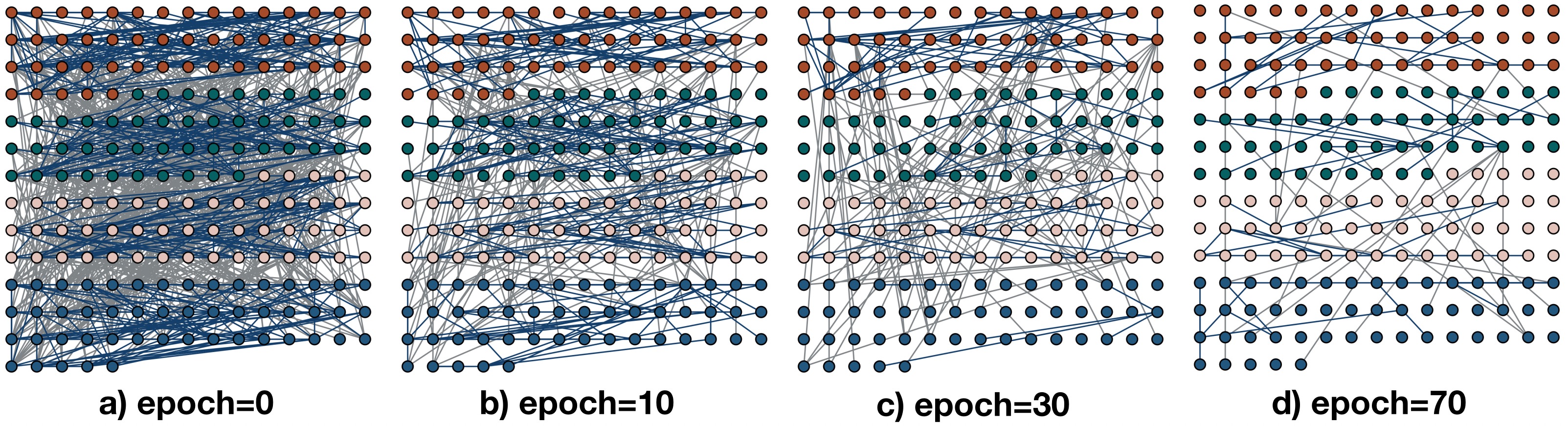}
% \vspace{-5pt}
\caption{The grid layout of read-overlap graph with the violating edges in the training of \texttt{NeurHap}.}
\label{fig5}
% \vspace{-10pt}
\end{figure}

Figure~\ref{fig5} shows the search process of \texttt{NeurHap} on the Sim-Potato-5X Sample 1 as an example. The sub-figure a) shows the grid layout of the initial coloring of the read-overlap graph violates significant number of conflicting edges (in grey) and consistent edges (in blue).  With the increasing number of epochs, the number of violating constraints (conflicting and consistent edges) decrease significantly.
% When epoch reaches 273, the number of violating constraints is 0 indicating an optimal solution.
%which infers that the model obtain a optimal solution.

\subsection{Experimental Analysis}

% \begin{wrapfigure}{r}{0.4\textwidth}
% % \vspace{-10pt}
%   \centerline{\includegraphics[width=0.4\textwidth]{figures/figure_06.jpg}}
% %   \vspace{-5pt}
%   \caption{Results of
%   \texttt{NeurHap}-search and \texttt{NeurHap} (\texttt{NeurHap}-search + \texttt{NeurHap}-refine) on all five datasets.}
%   \label{fig6}
% % \vspace{-10pt}
% \end{wrapfigure}

\begin{wrapfigure}{r}{0.4\textwidth}
\vspace{-10pt}
  \centerline{\includegraphics[width=0.4\textwidth]{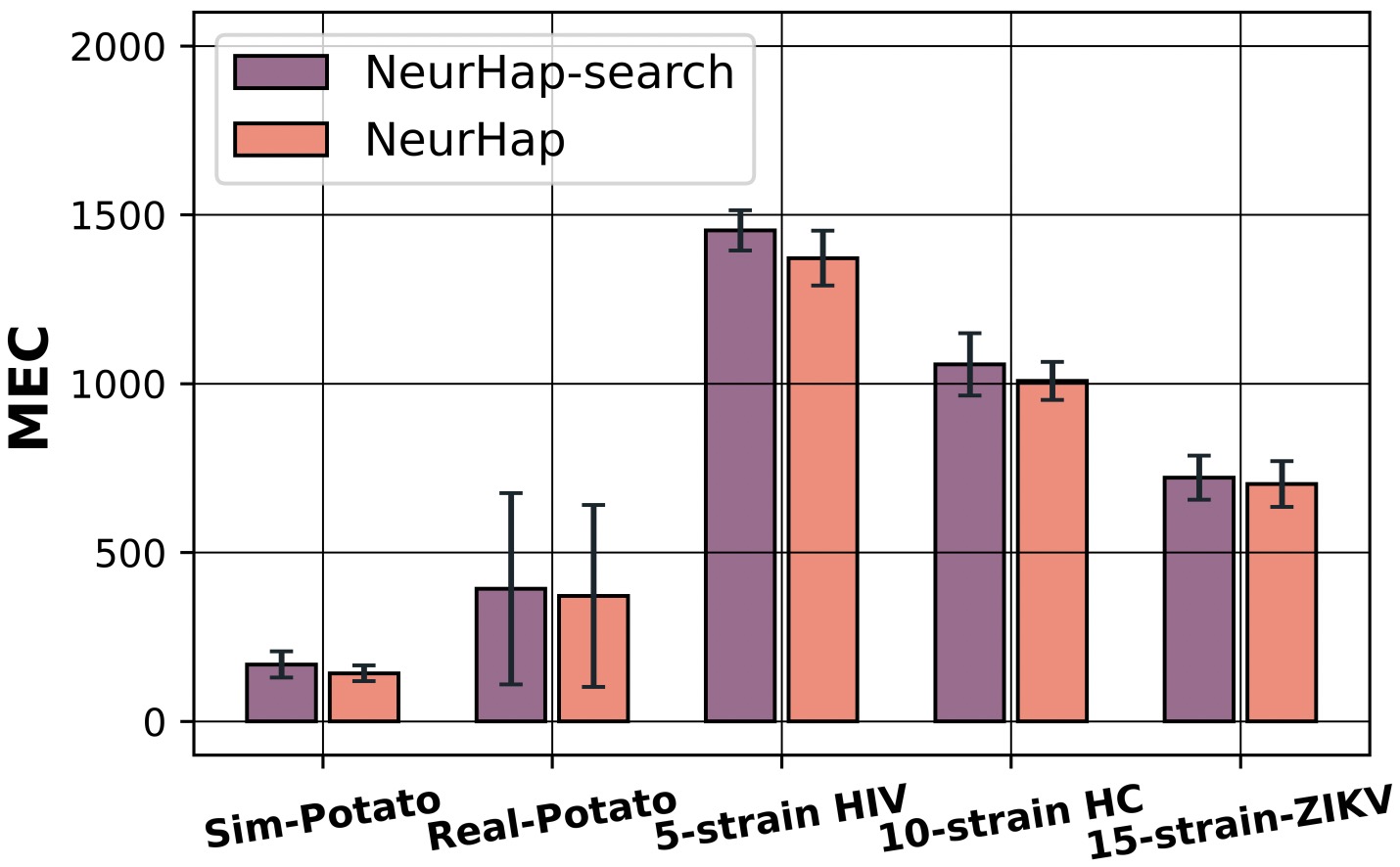}}
  \vspace{-5pt}
  \caption{Results of
  \texttt{NeurHap}-search and \texttt{NeurHap} (\texttt{NeurHap}-search + \texttt{NeurHap}-refine) on all five datasets.}
  \label{fig6}
\vspace{-10pt}
\end{wrapfigure}

\textbf{Ablation study.} To study the effectiveness of our proposed model, we 
 conduct an ablation study to 
examine the two algorithmic components in \texttt{NeurHap}, a graph neural network-based algorithm  \texttt{NeurHap}-search and a local combinatorial optimisation-based refinement algorithm  \texttt{NeurHap}-refine. Figure~\ref{fig6} shows that \texttt{NeurHap}-refine is able to further optimize the MEC score, e.g., the MEC scores for 5-strain HIV by \texttt{NeurHap}-search and \texttt{NeurHap} are 1453.5 and 1371.4 respectively, demonstrating the complementary effectiveness of global search and local refinement algorithms on phasing haplotypes.

\textbf{Parameter analysis \& Running time.} We investigate the importance of core parameters in model, including $p$ and $q$ for read-overlap graph, $\lambda$ for consistent constraints, $t$ for iterations, and $d$ for feature dimension. The detailed parameters analysis is listed in the Appendix A.3. We benchmark the running time of \texttt{NeurHap} against two deep learning baselines CAECseq and GAEseq on the Sim-Potato-Cov30 data. \texttt{NeurHap} achieves the lowest MEC score (142.0) compared with CAECseq (372.9) and GAEseq (496.9). 
The running time of \texttt{NeurHap} is 258 seconds which is faster than CAECseq (341 seconds). GAEseq is the slowest among the three and takes 492 seconds.

\section{Conclusion}
In this paper, we propose \texttt{NeurHap}, a graph representation learning approach to reconstruct haplotypes of polyploid species and viral quasispecies. We give a novel formulation of the haplotype phasing problem as a graph coloring problem.
%on a read-overlap graph which is constructed by computing the overlaps %ping MEC value 
%among different sequencing reads as well as criteria for pairwise consistency and conflicts. 
We design a message-passing based graph neural network search framework over a carefully constructed graph 
%read-overlap graph 
to assign colors (indicating haplotypes) to the reads, and a local refinement step to adjust colors to optimize MEC scores.
%\texttt{NeurHap}-search. We also design a local refinement algorithm, \texttt{NeurHap}-refine, to iteratively explore local color adjustments to further improve the final MEC scores that indicate good haplotype phasing.
%meanwhile satisfying conflict constraints. 
Extensive experiments on both simulated and real-world datasets demonstrate the effectiveness of our proposed \texttt{NeurHap} model on phasing haplotypes from polyploid species and viral quasispecies. 
A limitation of our method is in its ability to handle long reads. Massive long reads in polyploids leads to an even larger search space that may be addressed by extensions to our approach in future work.
%A shortcoming of our model is the extensibility of the massive numbers of long reads. Massive long reads with more polyploid species make the haplotype assembly more challenging which needs more search space. In the future, we plan to improve the effectiveness of combining graph neural networks and graph colouring over large-scale polyploid species.
Besides, NeurHap cannot automatically discover the number of haplotypes. This limitation will be addressed going forward.

\bibliographystyle{plainnat}
\bibliography{references}

\appendix
\section{Appendix}
% In this paper, we provide a unique formulation of the haplotype phasing problem as a graph coloring problem, and develop an algorithm based on graph representation learning and combinatorial optimization. 
% The proposed \texttt{NeurHap} consists of \texttt{NeurHap}-search, a graph neural network to learn vertex representations and color assignments followed by \texttt{NeurHap}-refine, a local refinement strategy to adjust colors and optimize MEC scores.

\subsection{The pseudocode for \texttt{NeurHap}-refine algorithm}
From the previous \texttt{NeurHap}-search step, we obtain an initial coloring assignment for vertices that satisfy the constraints of the read-overlap graph. However, it may exists multiple coloring assignments that satisfy all constraints. Therefore, we run an additional local refinement step to further optimise the MEC score. 
\texttt{NeurHap}-refine mainly searches for possible color adjustments of individual vertices given their associated conflicting and consistent constraints.
More specifically, if an individual vertex can be assigned a color different from its current color without violating any of associated conflicting constraints with the neighboring vertices, the color is changed if a better MEC score is obtained by the change.
The local refinement algorithm, \texttt{NeurHap}-refine, iteratively explores these possible color adjustments of individual vertices. 
The pseudocode for the \texttt{NeurHap}-refine is as follows:

{\centering
\begin{minipage}{.975\linewidth}
\begin{algorithm}[H]
\SetAlgoLined
\KwData{Read-overlap graph $\mathcal{G}$; number of polyploids $k$; initial color assignment $\mathcal{Y}$}
\KwResult{final color assignments $\mathcal{Y}^*$.}
Tag $\leftarrow$ True \ \ \ \ // \texttt{Initialize the iteration tag as True} \\
\While{Tag $==$ True}{
Tag $\leftarrow$ False \ \ \ \ // \texttt{Set the iteration tag as False} \\
\For{node $v\in\mathcal{V}$}{
$CN_{v}\leftarrow \{c(u)| (v,u) \in \mathcal{E}_{\neq}\}$\ \ \ \ \ \ //\ \texttt{Compute the set of colors from conflicting neighbors} \\
\For{$c' \notin CN(v) \mbox{~and~} c' \neq c(v) $}{
\ \ \ \ \ // \texttt{for every possible alternative color $c'$ for $v$} \\
$\mathcal{Y'} \leftarrow \mathcal{Y}_{c(v) \leftarrow c'} $ \\ \ \ \ \ \ //\ \texttt{$\mathcal{Y'}$} is derived by setting $c'$ as the color of $v$ in $\mathcal{Y}$\\
\If{MEC($\mathcal{Y'}$) < MEC($\mathcal{Y}$)}{
$\mathcal{Y} \leftarrow \mathcal{Y'}$ \ \ \ \ // \texttt{Move to a better coloring scheme} \\
Tag $\leftarrow$ True \ \ \ \ // \texttt{Set the iteration tag to be true} \\
}
}
}
}
$\mathcal{Y}^* \leftarrow \mathcal{Y}$ \ \ \ \ // \texttt{Output the final coloring scheme} \\
\caption{The Local Refinement Algorithm \texttt{NeurHap}-refine.}
\end{algorithm}
\end{minipage}
\par
}

\subsection{Implementation Details}
Two categories of datasets are used in the paper, \textit{Polyploid species} and \textit{ Viral Quasispecies}. \textit{Polyploid species} contains two datasets, Sim-Potato ($k=4$) and Real-Potato ($k=4$), which are downloaded from CAECseq~\citep{Ke2020CAECseq} and GAEseq~\citep{Ke2020GAEseq}. \textit{ Viral Quasispecies} contains three datasets, 5-strain HIV ($k=5$), 10-strain HCV ($k=10$), and 15-strain ZIKV ($k=15$), which are downloaded from SAVAGE~\citep{Baaijens2017DeNA}. It has two steps to generate the SNP matrix, i) Align reads to a reference genome and ii) Extract the matrix from the alignment. 

\textbf{i) Align Reads to Reference.}
BWA-MEM~\citep{Li2013AligningSR} is used to align reads to the reference genome. The detailed command is (take the 15-strain ZIKV as an example):

\texttt{\$ ./bwa index 15-strain-ZIKV.fasta}\\
\texttt{\$ ./bwa mem 15-strain-ZIKV.fasta forward.fastq reverse.fastq > 15-strain-ZIKV.sam}

\textbf{ii) Extract the SNP Matrix.}
We use the same tool described in CAECseq and GAEseq~\citep{Ke2020CAECseq,Ke2020GAEseq} to derive the SNP matrix from the above alignment to ensure a fair comparison. The default parameters are used in the configure file which is same with CAECseq and GAEseq. The detailed command is:

\texttt{\$ ./ExtractMatrix config}

For all five datasts, we randomly generate 10 samples. The detailed number of reads and SNPs for Real-Potato, 5-strain HIV, 10-strain HCV, and 15-strain ZIKV are listed in the paper. For Semi-Potato, sequencing coverage is varied from 5X to 30X. We have 40 sub-datasets in Semi-Potato. The read numbers range from approximately 200 to 1200 and the number of SNPs vary from 200 to 400. 

\textbf{Read-overlap Graph.} After obtaining the SNP matrix, we build the consistent and conflicting edges between pairs of reads (i.e., pairs of rows in the SNP matrix). Two parameters are introduced in this step to the construct read-overlap graph, $p$ and $q$. 
% Two reads are overlapping if they overlap in at least $p$ SNP positions otherwise non-overlapping. Two overlapping reads $R_i$ and $R_j$ are consistent if they have the same alleles over all SNP positions (i.e., $HD(R_i, R_j) = 0$), and are in conflict if they differ on at least $q$ SNP positions (i.e., $HD(R_i
% , R_j) \geq q$), where $HD(R_i, R_j)$ represents the Hamming distance between two overlapping reads in the read-overlap graph.
Two overlapping reads $R_i$ and $R_j$ are \emph{consistent} if they have the same alleles over all SNP positions meanwhile the length of overlapping is larger than $p$ (i.e., $HD(R_i,R_j) = 0$), and are in \emph{conflict} if they differ on at least $q$ SNP positions (i.e., $HD(R_i,R_j) \geq q$), where $HD(R_i,R_j)$ represents the Hamming distance  between two overlapping reads in the read-overlap graph.
We adjust two thresholds according different datasets from 2 to 6, and we also evaluate the effect of two parameters for the \texttt{NeurHap} model. 

% We submit the codes for \texttt{NeurHap} in the supplementary material. 
You can simply run the following to reproduce the experimental results (we take the Sim-Potato-Cov5 Sample 1 as an example).

\texttt{\$ python main.py -e 2000 -t 10 -f 32 -k 4 -r 1e-3 -p 6 -q 2 -l 0.01 -d Semi-Potato -s Sample1}

where parameter $\texttt{-e}$ represents the number of epoch, $\texttt{-t}$ is the number of the iteration, $\texttt{-f}$ is the dimension of the embedding, $\texttt{-k}$ is the number of haplotypes or ploids, $\texttt{-r}$ is the learning rate, $\texttt{-l}$ denotes the $\lambda$. Parameters $\texttt{-d}$ and $\texttt{-s}$ are used to select the corresponding data and sample. 
% All codes, data and experimental settings for \texttt{NeurHap} model will be released after the double-blind review. 
The source code of NeurHap is freely available at \url{https://github.com/xuehansheng/NeurHap}.

\textbf{Running environment.} \texttt{NeurHap} is implemented in Python 3.6 and Pytorch 1.8 using the Linux server with 6 Intel(R) Core(TM) i7-7800X CPU @ 3.50 GHz, 96GB RAM and 2 NVIDIA RTX A6000 with 48GB memory. 
% The NeurHap model is freely available at \url{https://github.com/xuehansheng/NeurHap}.

% 2 NVIDIA TITAN Xp 12 GB. 

\subsection{Experimental Analysis}
\begin{figure}[h]
\centering
\includegraphics[width=0.7\columnwidth]{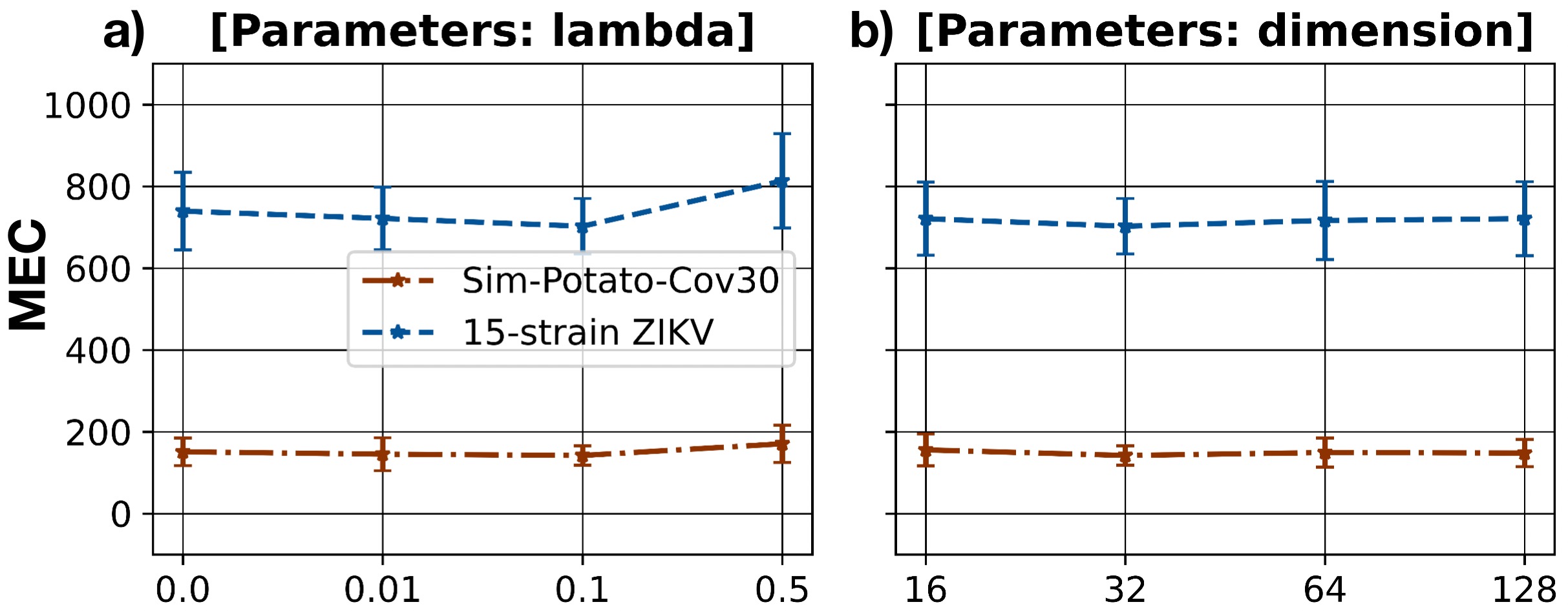}
\caption{Parameters analysis of the \texttt{NeurHap} model ($\lambda$ and $d$).}
\label{figS1}
\end{figure}

\textbf{Parameters Analysis.} In this section, we investigate the importance of core parameters in model, including $p$ and $q$ for read-overlap graph, $\lambda$ for consistent constraints, $t$ for iterations, and $d$ for feature dimension. Figure~\ref{figS1} b) shows that our proposed \texttt{NeurHap} is robust to the dimension of latent embedding $d$. In Figure~\ref{figS1} a), the MEC scores for \texttt{NeurHap} with $\lambda$ varying from 0.0 to 0.1 do not change too much and relatively stable. However, if we choose $\lambda$ as 0.5, the performance of \texttt{NeurHap} being worse. We vary $\lambda$ from 0.0 to 0.1 for \texttt{NeurHap}.

\begin{figure}[h]
\centering
\includegraphics[width=0.85\columnwidth]{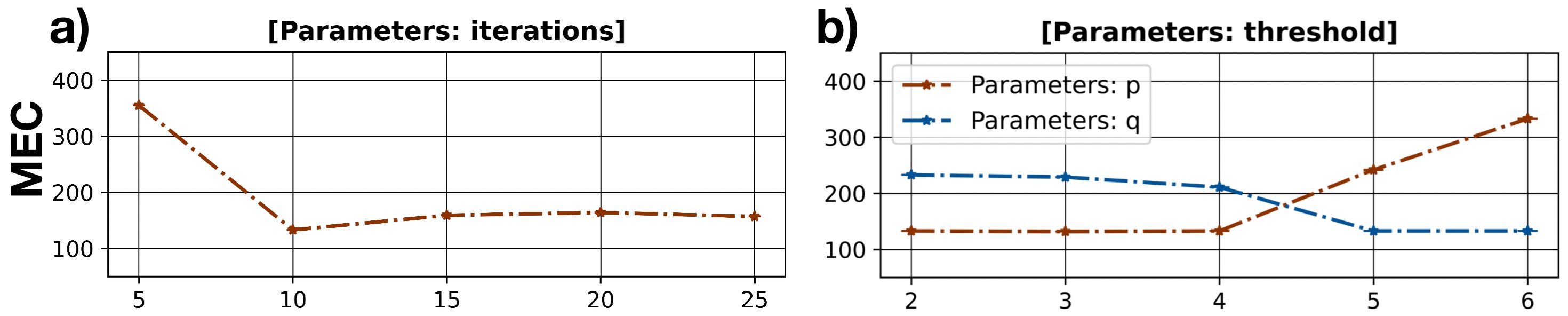}
\caption{Parameters analysis of the \texttt{NeurHap} model ($t$, $p$, and $q$).}
\label{figS2}
\end{figure}

Next, we investigate the effects of parameters $t$, $p$,  and $q$ (take the Sim-Potato-Cov30X Sample 1 as an example). We vary iteration $t$ from 5 to 25 and the results are shown in Figure~\ref{figS2} a). When the iteration $t=10$, \texttt{NeurHap} achieves the best performance on the Sim-Potato-Cov30X Sample 1 dataset. When the iteration $t\geq10$, the MEC score of the \texttt{NeurHap} is relatively stable. 
In Figure~\ref{figS2} b), When parameters $q=4$ and $p=5$, the \texttt{NeurHap} achieves the best performance. 
If the parameter $p<5$ ($q$ is fixed to 4), the MEC score of the \texttt{NeurHap} is high because the constructed consistent edges are not confident and they contain several mistaken consistent edges. When the parameter $q>4$ (the $p$ is fixed to 5), the number of extracted conflicting edges is few (the read-overlap graph is sparse) which is not good to optimise the MEC score. 
% Thus, when parameters $p<5$ or $q>4$ in the Sim-Potato-Cov30X Sample 1 dataset, the MEC score of the \texttt{NeurHap} is high. 

% \begin{wrapfigure}{r}{0.3\textwidth}
% \vspace{-15pt}
%   \centerline{\includegraphics[width=0.3\textwidth]{figures/figure_08.jpg}}
% \vspace{-7pt}
%   \caption{The running time.}
%   \label{fig8}
% \vspace{-20pt}
% \end{wrapfigure}

\textbf{Running Time.} 
We benchmark the running time of \texttt{NeurHap} against two deep learning baselines CAECseq and GAEseq on the Sim-Potato-Cov30 Sample 1 data. NeurHap achieves the lowest MEC score (142.0) compared with CAECseq (372.9) and GAEseq (496.9). The running time of NeurHap is 258 seconds which is faster than CAECseq (341 seconds). GAEseq is the slowest among the three and takes 492 seconds. 

\begin{figure}[h]
  \centerline{\includegraphics[width=0.3\textwidth]{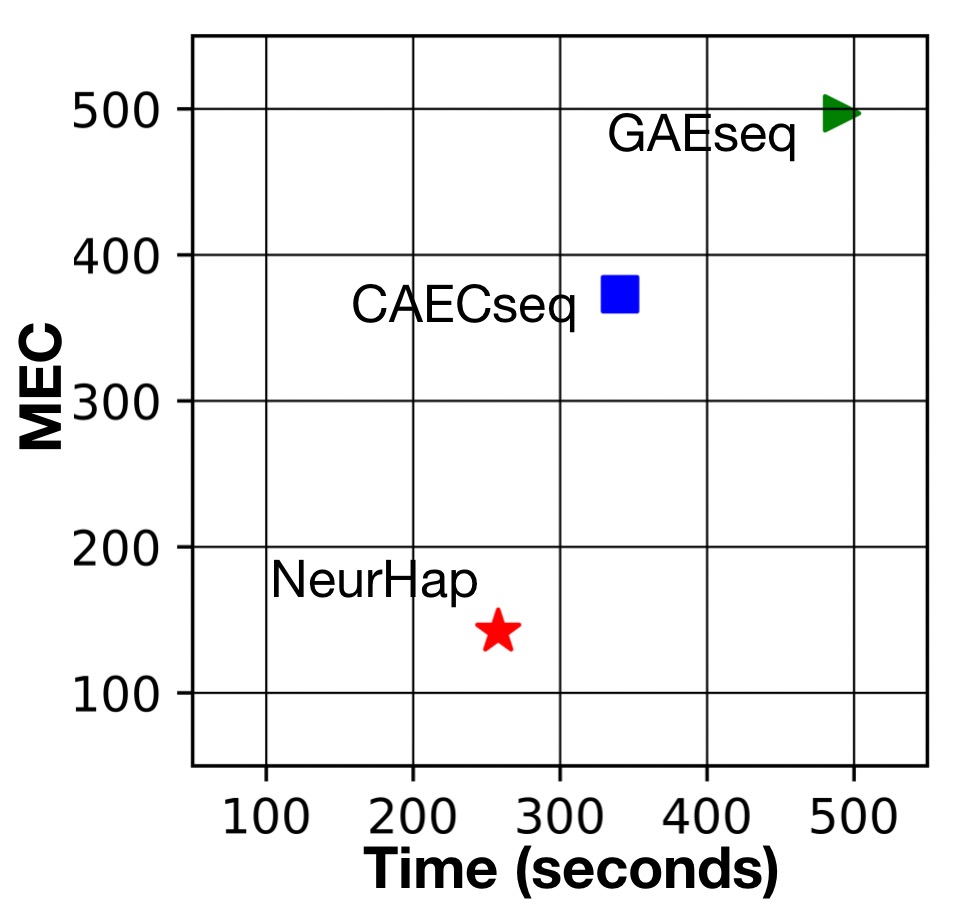}}
  \caption{The running time of \texttt{NeurHap}, CAECseq and GAEseq.}
  \label{fig8}
\end{figure}

\subsection{Additional Experiment}
\textbf{Average MEC on Semi-Potato.} In the experimental part,  we select the lowest MEC score as the final results after running experiment 10 times which is same as previous SOTA baselines~\cite{Ke2020CAECseq,Ke2020GAEseq}. Here, we also report the average MEC score after running all algorithms on Semi-Potato 10 times (see Table~\ref{tab:avg}). \texttt{NeurHap} still outperforms other SOTA baselines.

\begin{table}[!htb]
 \caption{Performance comparison on Sim-Potato (Tetraploid, $k=4$).}
  \centering 
  \footnotesize
  \begin{tabular}{c|c|c|c|c|c}
    \toprule
    \textbf{Polyploids} & \textbf{Model} & \textbf{\#Cov 5X} & \textbf{\#Cov 10X} & \textbf{\#Cov 20X} & \textbf{\#Cov 30X} \\
    \midrule
     \multirow{5}{*}{Tetraploid} & H-PoP & 429.0\scriptsize$\pm$64.1 & 933.9\scriptsize$\pm$103.6 & 1782.2\scriptsize$\pm$161.8 & 2826.9\scriptsize$\pm$180.7 \\
     \multirow{5}{*}{(\textit{k=4})} & AltHap & 610.9\scriptsize$\pm$259.3 & 722.3\scriptsize$\pm$179.1 & 649.3\scriptsize$\pm$369.4 & 1148.2\scriptsize$\pm$509.9 \\
     & GAEseq & 225.1\scriptsize$\pm$17.7 & 391.2\scriptsize$\pm$45.5 & 610.4\scriptsize$\pm$97.3 & 811.8\scriptsize$\pm$131.8 \\
     & CAECseq & \underline{160.5\scriptsize$\pm$25.9} & \underline{266.0\scriptsize$\pm$43.3} & \underline{466.5\scriptsize$\pm$89.0} & \underline{629.5\scriptsize$\pm$160.0} \\
    \cmidrule{2-6}
     & NeurHap & \textbf{37.5\scriptsize$\pm$5.5} & \textbf{62.8\scriptsize$\pm$7.5} & \textbf{113.2\scriptsize$\pm$19.8} & \textbf{166.3\scriptsize$\pm$26.7} \\
    \bottomrule
  \end{tabular}
  \label{tab:avg}
\end{table}

\textbf{Benchmark against Graph Coloring.} We also benchmark \texttt{NeurHap} against two graph coloring algorithms, including Greedy~\citep{Daniel1979Greedy} and RUN-CSP~\citep{Toenshoff2019RUNCSPUL}. We implement graph coloring algorithms on the read-overlap graphs which only contain conflicting edges because those methods cannot address the consistent edges. In Table~\ref{tab:graphcolor}, \texttt{NeurHap} significantly outperforms graph coloring algorithms.

\begin{table}[!htb]
 \caption{Performance comparison on Real-Potato ($k=4$) and 5-strain HIV ($k=5$).}
  \centering 
  \footnotesize
  \setlength{\tabcolsep}{1.1mm}
  \begin{tabular}{c|c|ccccccccccc}
    \toprule
    \textbf{Data} & \textbf{Model} & \# 1 & \# 2 & \# 3 & \# 4 & \# 5 & \# 6 & \# 7 & \# 8 & \# 9 & \# 10 & \# Avg. \\
    \midrule
    \multirow{3}{*}{Real-Potato} & Greedy & 296 & 458 & 162 & \underline{3} & 239 & 1014 & 679 & 602 & 694 & 906 & 505.3\scriptsize$\pm$330.3  \\
    & RUN-CSP & \underline{186} & \underline{358} & \underline{107} & \textbf{1} & \underline{185} & \underline{890} & \underline{553} & \underline{492} & \underline{647} & \underline{767} & \underline{418.6\scriptsize$\pm$298.7} \\
    % \cmidrule{3-13}
    & NeurHap & \textbf{178} & \textbf{343} & \textbf{93} & \textbf{1} & \textbf{163} & \textbf{857} & \textbf{499} & \textbf{384} & \textbf{561} & \textbf{632} & \textbf{371.6\scriptsize$\pm$268.9} \\
    \midrule
    \multirow{3}{*}{5-strain HIV} & Greedy & 3974 & 3791 & 3633 & 3819 & 4251 & 3472 & 3137 & 3241 & 3326 & 3476 & 3612.0\scriptsize$\pm$349.3  \\
    & RUN-CSP & \underline{2226} & \underline{2375} & \underline{2175} & \underline{2408} & \underline{2192} & \underline{2748} & \underline{2449} & \underline{2614} & \underline{2312} & \underline{2567} & \underline{2406.6\scriptsize$\pm$191.3} \\
    % \cmidrule{3-13}
     & NeurHap & \textbf{1307} & \textbf{1525} & \textbf{1385} & \textbf{1265} & \textbf{1410} & \textbf{1382} & \textbf{1393} & \textbf{1323} & \textbf{1274} & \textbf{1450} & \textbf{1371.4\scriptsize$\pm$81.2}\\
    \bottomrule
  \end{tabular}
  \label{tab:graphcolor}
\end{table}

\textbf{MEC score v.s. Violating Constraints.} 
While Eqn. 2 aims to minimize the sum of hamming distances between each read $\mathcal{R}_j$ and the haplotype $\mathcal{H}_i$ that is drawn from $\mathcal{R}_j$, Eqn. 3 aims to minimize the divergence between pairs of reads (as $P(v_i)$ and $P(v_j)$) that are drawn from the same haplotype and maximize the divergence between pairs of reads if they are drawn from different haplotypes. Moreover, the hamming distances in Eqn.2 have been used implicitly to derive in Eqn.3. 
In an ideal case, if all pairs of conflicting reads are assigned into different haplotypes (i.e., different colors) and all pairs of consistent reads are assigned into the same haplotypes (i.e., the same color), each cluster will only contain consistent reads and thus the MEC score in Eqn.2 will be minimized to be 0. 
In non-ideal cases such as Sim-Potato Cov-5X Sample 1 datasets, the following Figure~\ref{fig9} shows that the objective function to be minimized in Eqn.2 (i.e., MEC) correlates well with the objective function to be minimized in Eqn.3, which demonstrates the effectiveness of NeurHap for minimizing the MEC through optimizing Eqn.3.

\begin{figure}[h]
  \centerline{\includegraphics[width=0.99\textwidth]{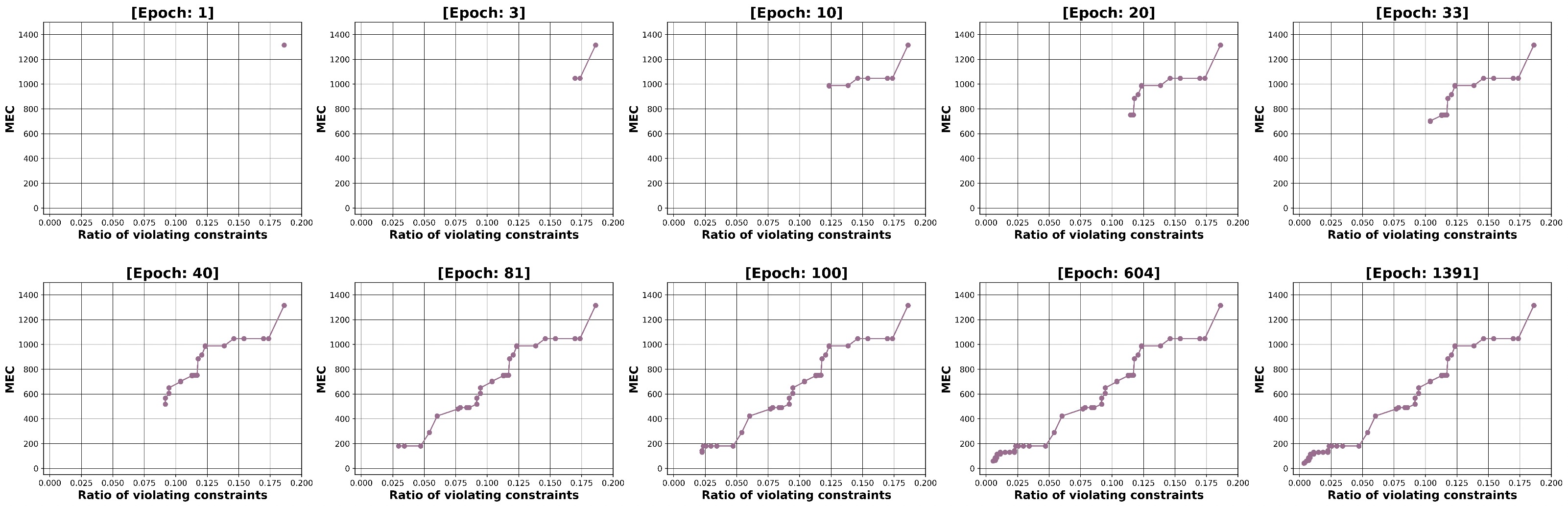}}
  \caption{The MEC score v.s. Violating Constraints in the training process of \texttt{NeurHap}.}
  \label{fig9}
\end{figure}

\textbf{Scalability.}
To evaluate the scalability of the NeurHap, we incrementally combine the samples in Real-Potato dataset and summarise the results in Table~\ref{tab:scalability}. It is observed that the running time of NeurHap is roughly linearly correlated with the number of total edges (conflict edges + consistent edges). On the other hand, diploid haplotype assembly remains challenging for reconstructing chromosome-level haplotypes, especially for large eukaryotic genomes with complex repeats. Similar to CAECseq and GAEseq, NeurHap has also focused on short-read datasets on gene regions because complex repeats in the intergenic regions along the chromosome make it impossible to reconstruct continuous haplotypes reliably. 
% As NeurHap has demonstrated its potential scalability, in future we will investigate whether NeurHap can be applied to long-read datasets to reconstruct chromosome-level haplotypes.

\begin{table}[!htb]
 \caption{Performance comparison on cumulative Real-Potato dataset.}
  \centering 
  \footnotesize
  \setlength{\tabcolsep}{1.1mm}
  \begin{tabular}{c|c|ccccccccccc}
    \toprule
    \multicolumn{2}{c|}{\textbf{Samples}} & \# 1 & \# 2 & \# 3 & \# 4 & \# 5 & \# 6 & \# 7 & \# 8 & \# 9 & \# 10 \\
    \midrule
    \multicolumn{2}{c|}{Reads} & 240 & 629 & 903 & 1018 & 1159 & 1557 & 1852 & 2136 & 2625 & 3074 \\
    \multicolumn{2}{c|}{SNPs} & 295 & 533 & 616 & 639 & 815 & 1013 & 1469 & 1893 & 2129 & 2539 \\
    \multicolumn{2}{c|}{Conflict} & 3351 & 6380 & 13433 & 14208 & 16207 & 34288 & 39368 & 42811 & 51358 & 59621 \\
    \multicolumn{2}{c|}{Consistent} & 966 & 1514 & 3323 & 3537 & 3908 & 5747 & 6470 & 6977 & 7985 & 9329 \\
    \midrule
    \multirow{2}{*}{CAECseq} & MEC & 229 & 786 & 910 & 985 & 1282 & 1997 & 2584 & 3018 & 3914 & 4524 \\
    & time & 243s & 283s & 302s & 310s & 414s & 586s & 798s & 1188s & 1991s & 2774s \\
    \midrule
    NeurHap & MEC & 183 & 559 & 671 & 692 & 888 & 1802 & 2305 & 2667 & 3316 & 3992 \\
    -search & time & 28s & 38s & 52s & 53s & 63s & 99s & 128s & 157s & 200s & 253s \\
    \bottomrule
  \end{tabular}
  \label{tab:scalability}
\end{table}

Besides, we applied NeurHap on a chromosome-level dataset for Chromosome 22 of the human genome to validate the scalability of NeurHap. Specifically, we downloaded publicly available alignment files for the Human Genome NA12878 (from \url{http://s3.amazonaws.com/nanopore-human-wgs/NA12878-Albacore2.1.sorted.bam}) and combined them with the set of heterozygous SNPs on Chromosome 22 of the human genome (derived from~\citep{Duitama2012FosmidbasedWG}) to build the input alignment matrix (following the same procedure introduced in HapCUT2~\citep{Edge2017HapCUT2RA}). This constructed matrix contains 129,338 long reads and 22,792 SNPs. NeurHap took 734 seconds and around 13G memory to reconstruct two chromosome-level haplotypes with a MEC score of 23,114. As Chromosome 22 is about 1.6\% of the whole human genome and 20\% of the largest chromosome (Chromosome 1) in the human genome, we estimate (optimistically) that phasing all the chromosomes in the human genome will take about 12 hours with a peak memory of 65G. Note that CAECseq and GAEseq are both out of the memory when they were applied to this chromosome-level dataset on the NVIDIA RTX A6000 with 48GB memory.

\end{document}